\documentclass[twocolumn, superscriptaddress]{revtex4-1}
\usepackage{amsmath}
\usepackage{amsfonts}
\usepackage{amssymb}
\usepackage{xspace}
\usepackage[final]{graphicx}
\usepackage[svgnames]{xcolor}
\usepackage[USenglish]{babel}
\usepackage{siunitx}
\usepackage[pdftex,colorlinks=true,citecolor=blue]{hyperref}

\renewcommand{\vec}{\mathbf}
\newcommand{\fs}[1]{\SI{#1}{\femto\second}}
\newcommand{\um}[1]{\SI{#1}{\micro\metre}}
\newcommand{\nm}[1]{\SI{#1}{\nano\metre}}
\newcommand{\ev}[1]{\SI{#1}{\electronvolt}}
\newcommand{\tief}[1]{\ensuremath{_\text{\tiny #1}}}
\newcommand{\kspp}{\ensuremath{k\tief{hspp}}\xspace}
\newcommand{\ie}{\textsl{i.e.}\xspace}

\renewcommand{\Re}{\operatorname{Re}}
\renewcommand{\Im}{\operatorname{Im}}
\newcommand{\e}{\operatorname{e}}

\begin{document}
\title{Characterization of a circular optical nanoantenna by nonlinear photoemission electron microscopy}
\author{Thomas Kaiser}
\author{Matthias Falkner}
\affiliation{Institute of Applied Physics, Abbe Center of Photonics, Friedrich-Schiller-Universit\"at Jena, Max-Wien-Platz 1, 07743 Jena, Germany}
\author{Jing Qi}
\affiliation{Institute of Condensed Matter Theory and Solid State Optics, Abbe Center of Photonics, Friedrich-Schiller-Universit\"at Jena, Max-Wien-Platz 1, 07743 Jena, Germany}
\author{Angela Klein}
\author{Michael Steinert}
\author{Christoph Menzel}
\affiliation{Institute of Applied Physics, Abbe Center of Photonics, Friedrich-Schiller-Universit\"at Jena, Max-Wien-Platz 1, 07743 Jena, Germany}
\author{Carsten Rockstuhl}
\affiliation{Institute of Theoretical Solid State Physics and Institute of Nanotechnology, Karlsruhe Institute of Technology, 76131 Karlsruhe, Germany}
\author{Thomas Pertsch}
\affiliation{Institute of Applied Physics, Abbe Center of Photonics, Friedrich-Schiller-Universit\"at Jena, Max-Wien-Platz 1, 07743 Jena, Germany}

\date{14.08.2015}

\begin{abstract}
We report on the investigation of an advanced circular plasmonic nanoantenna under ultrafast excitation using nonlinear photoemission electron microscopy (PEEM) under near-normal incidence.
The circular nanoantenna is enhanced in its performance by a supporting grating and milled out from a  gold film.
The considered antenna shows a sophisticated physical resonance behavior that is ideal to demonstrate the possibilities of PEEM for the experimental investigations of plasmonic effects on the nanoscale.
Field profiles of the antenna resonance for both possible linear polarizations of the incident field are measured with high spatial resolution.
In addition, outward propagating Hankel plasmons, which are also excited by the structure, are measured and analyzed.
We compare our findings to measurements of an isolated plasmonic nanodisc resonator and scanning near-field optical microscopy (SNOM) measurements of both structures.
All results are in very good agreement with numerical simulations as well as analytial models that are also discussed in our paper.
\end{abstract}

\maketitle

\section{Introduction}
\label{introduction}

Optical nanoantennas provide a possibility to convert light between the far-field and nanoscale dimensions and constitute building-blocks of utmost importance in nanooptics \cite{Bharadwaj2009,Novotny2011,Biagioni2012,Krasnok2013}.
They are typically created from sub-wavelength plas\-mo\-nic resonators.
A variety of different design approaches has emerged, from nanowire \cite{Dorfmuller2010} or bow-tie antennas \cite{Schuck2005} to sophisticated geometries where the emission characteristic and resonance strength can be tailored by design \cite{Qi2015}.
The short response time of the involved plasmonic excitations allows to enter the realm of ultrafast light control at the nanoscale with these structures, opening entirely new research perspectives for physics, chemistry, biology, and material science \cite{Stockman2007,Aeschlimann2007}.

The success of such approaches critically depends on the possibility to investigate the structures experimentally on the nanoscale.
Scanning near-field optical microcopy (SNOM) has become an established and accepted method for this purpose \cite{Rotenberg2014,Wulf2014}.
However, the presence of a physical probe, which disturbs the optical functionality of the structure and limits the spatial resolution, as well as the scanning nature of the method prompted the development of alternative experimental techniques \cite{Douillard2013}.

Photoemission electron microscopy (PEEM) has become a major tool for this purpose in recent years when combined with ultrafast light sources \cite{Fecher2002,Schmidt2002,Cinchetti2005, Stockman2007,Aeschlimann2007,MeyerzuHeringdorf2007,Kubo2007,Chelaru2007,Chew2012,Gong2015}.
The interference of surface plasmons and the incident light pulse allows the direct visualization of the plasmon field in the nonlinear regime, \ie when multiple photons are required to excite one photoelectron.
Since most PEEM experiments are done under a grazing angle of incidence, wide beating patterns and Moir\'e-like images are observed when imaging the straight propagation of surface plasmons excited by slits, trenches, or other straight edges \cite{Kubo2007,Zhang2011,Buckanie2013,Lemke2013b,Lemke2014b,Gong2015}.
The interference from structures with circular symmetry like nanoholes, however, show wide parabolic ring-like patterns \cite{Lemke2014,Gong2014}.
This limitation on the achievable image quality is mitigated when normal or near-normal incidence is used \cite{Kahl2014}.

The aforementioned publications concentrated on the investigation of simple propagating surface plasmon geometries or hot-spots generated by nanoparticles.
The study of more complex systems with a certain nanooptical functionality using nonlinear PEEM has so far been limited to the investigation of bow-tie nanoantennas \cite{Melchior2011}, plasmonic focusing devices \cite{Lemke2013}, nanoparticle-on-plane geometries \cite{Schertz2012}, or cross resonant optical antennas \cite{Klaer2015}.

In this work, we wish to investigate the behavior of a circular optical nanoantenna under ultrafast excitation using nonlinear PEEM at near-normal incidence. 
The resonance of the antenna was enhanced by a properly designed surrounding grating \cite{Qi2015}.
We find a pronounced enhancement of the photoelectron yield both for TE and TM polarization of the incident light.
Our results are compared to PEEM measurements on an isolated nanodisc antenna as well as SNOM measurements on both structures.
PEEM is found to be able to resolve the field-profile of the antenna modes with much better spatial resolution than SNOM.
The ring-like structure also supports the excitation of outward propagating Hankel plasmons on the surrounding gold film \cite{Nerkararyan2010}, which we also observe for both polarizations.
The measurements are accompanied by rigorous numerical simulations as well as analytical modeling of the interference of Hankel plasmons and the illumination under ultrafast excitation conditions, which explain the experimental findings with very good agreement.

\section{Circular optical nanoantenna with enhanced resonance}
\label{circ_antenna}

We start by introducing the physical working principle of the considered structure.
A plasmonic nanoantenna generally consists of a small metallic nanostructure that can sustain a plasmonic mode which is able to oscillate in resonance.
The resonance frequencies and other properties such as the field profile or emissivity of the nanoantenna are determined by the size, shape and material of the structure and can be tuned to fit a particular intended functionality \cite{Novotny2011}.

\subsection{Central gold nanodisc}
\begin{figure}
	\centering
	\includegraphics[width=8cm]{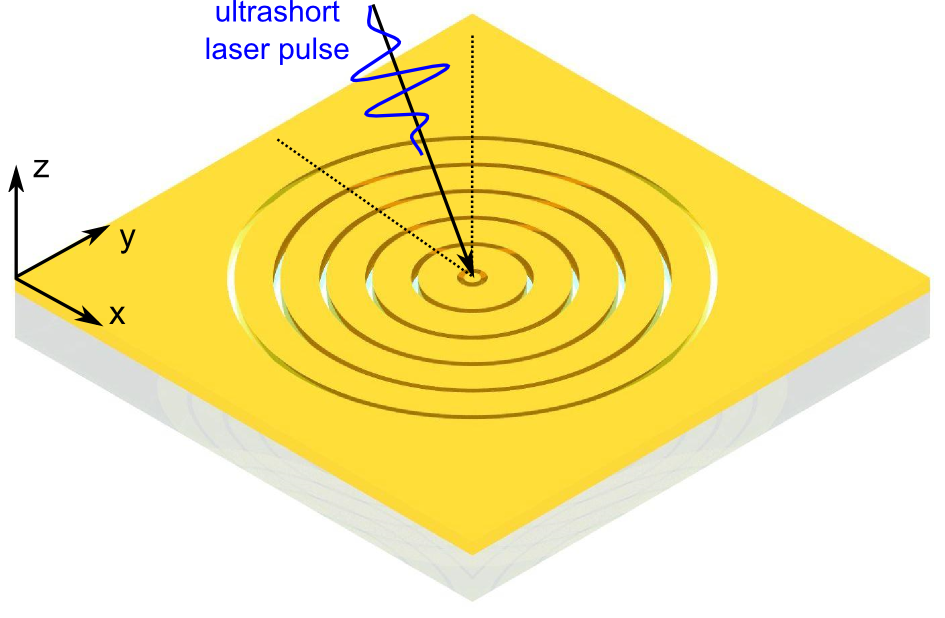}
	\caption{Scheme of the used sample and experimental configuration. A central nanodisc is surrounded by a circular grating with period $ p $. The grating helps enhancing the resonance in the central disc. The entire structure is surrounded by a gold film and the configuration is probed by an ultrashort laser pulse.}
	\label{fig_design}
\end{figure}

Our basic design is shown in Fig.~\ref{fig_design}.
The central resonant element is a circular gold nanodisc.
For such a system with circular symmetry, in- and outward propagating Hankel surface plasmon polaritons (HSPPs) are the fundamental solutions for the electromagnetic surface modes. 
At a fixed frequency, the electric field component perpendicular to the surface is described as \cite{Filter2012}
\begin{equation}
E_z(\vec r, \omega) \propto H_l^{(1,2)}[\kspp(\omega) r] \e^{il\varphi} ,
\end{equation}
where $ r = [x^2 + y^2]^\frac 1 2  $ is the radial coordinate and $ \varphi $ denotes the angular coordinate in the $x$-$y$-plane.The propagation constant of the plasmon is denoted as \kspp and reads as \cite{Maier2007, Filter2012}
\begin{equation}
\kspp(\omega) = k_0 \left[ \frac {\epsilon_m(\omega)} {\epsilon_m(\omega) + 1} \right]^{\frac 1 2} ,
\end{equation}
where $ \epsilon_m $ is the permitivity of gold.
Vacuum was assumed as cladding.
The $ \omega $-dependence of \kspp is not explicitly noted hereafter.
The Hankel functions of first or second kind and order $ l $ are denoted by $  H_l^{(1,2)} $, where $ l $ is an integer.
With a mathematical approximation for the Hankel function \cite{Abramowitz1964}, an expression for the total electric field of the HSPP is found 
\begin{equation} \label{eq:hspp}
\vec E\tief{hspp}(\vec r, \omega) = 
\left[\kspp r\right]^{-\frac 1 2} \e^{i\kspp r} \e^{il\varphi} 
\left( \vec e_z + i \frac{\kappa}{\kspp} \vec e_r \right), 
\end{equation}
where $ \kappa = [\kspp^2 - k_0^2\epsilon_m]^\frac 1 2 $.
Vectors $ \vec e_i $ represent unit vectors in $ i $-direction.
We have neglected an additional term 
\begin{equation}
\frac{\kappa}{\kspp^2 r}  \left(\frac 1 2 \vec e_r + il\vec e_\varphi\right)
\end{equation}
since it is quickly damped away with growing distance from the center (first several oscillations).
We may thus call \eqref{eq:hspp} a ''far-field'' approximation of the Hankel plasmon. 

The localized resonance of the nanodisc can be understood by regarding the structure as a  modal Fabry-Perot resonator \cite{Hasan2011}.
The outward-propagating HSPPs are partially reflected back into inward-propagating modes at the circumference.
This modal reflectivity can be tailored by proper design of the mode mismatch at the circumference \cite{Kaiser2013a}.
The resonance condition is a constructive interference and reads 
\begin{equation}
\kspp R + \Phi = \nu \cdot \pi,
\end{equation}
where $ R $ is the radius of the disc and $\Phi$ is the phase acquired upon reflection at the circumference. The radial mode number is denoted as $ \nu $ and needs to be an integer.
We use $ \nu = 1 $ in our case.

\subsection{Surrounding ring grating for enhanced resonance}
The resonance strength of a nanoantenna can be increased significantly when the central element is enclosed by a periodic structure \cite{Qi2014, Qi2015}.
The configuration acts as a Bragg reflector if concentric rings are used in our case to enclose the central nanodisc.
One can tune the periodicity such that the condition for resonant coupling of outward- into inward-propagating HSPPs is met.
This Bragg condition for HSPPs reads as
\begin{equation} \label{eq:bragg}
\kspp p = m \cdot \pi,
\end{equation}
where $ m $ is an integer denoting the band index and $ p $ is the period of the grating.
The grating increases the resonance strength in the central disc by enhancing the partial reflection of outward-propagating HSPPs at its circumference.
This increase in modal reflectivity enhances the quality factor of the resonator and leads to a better performance of the nanoantenna.

A second possibility to employ the grating considers the excitation of the antenna by incident light.
The bare disc can be excited from free-space illumination in the quasi-static regime thanks to its subwavelength size \cite{Bharadwaj2009}.
Such an approach does not require phase matching to the Hankel modes and can be used to excite the disc at any wavelength within the bandwidth of its localized plasmonic resonance.
However, it is also possible to employ the grating to couple incident light resonantly into Hankel plasmons \cite{Qi2015}.
The prerequisite for this is known as the Bloch condition and reads as
\begin{equation} \label{eq:bloch}
k_0\sin\theta + n\cdot \frac {2\pi} p = \kspp,
\end{equation}
where $ n $ is an integer.
The angle of incidence is denoted by $ \theta $.

Our design approach originates from the simple observation that \eqref{eq:bragg} and \eqref{eq:bloch} can be fulfilled simultaneously for certain combinations of $ \theta $, $m$, and $n$ if
\begin{equation}
\kspp \left(1 - \frac {2n} m \right) - k_0 \sin \theta = 0.
\end{equation}
Since we want to excite the structure from near-normal incidence ($\theta \approx \SI{0}{\degree}$), we use a grating corresponding to $ m=2 $ and $ n=1 $.
The period in our structure was thus chosen to be equal to the HSPP wavelength $ p = 2\pi / \kspp = \lambda\tief{hspp}$.

In this way, the circular grating works as a three-mode-coupler.
The occurrence of a Bragg resonance facilitates the interaction of the inward- and out\-ward-propagating Hankel mode which manifests itself by the formation of a bandgap.
The HSPPs become evanescent in the grating region and a coupling between the two otherwise orthogonal modes is possible within the penetration depth of the modes into the grating.
While the usually used first order Bragg resonance is located at the band edges, the second order Bragg resonance appears at the $ \Gamma $-point, \ie for a vanishing in-plane quasi-wavevector.
This facilitates the interaction of both the in- and outward HSPP mode with near-normal incident light.
The coupling to inward propagating HSPPs will further enhance the resonance of the nanoantenna whereas the excited outward propagating HSPP will be visible on the gold film surrounding the structure.

\section{PEEM measurements}
\label{peem_setup}

\subsection{Sample preparation and experimental setup}
We investigated the nanoantenna structure by means of multiphoton photoemission electron microscopy (PEEM).
Samples for the ring-enhanced nanoantenna structure and also the bare nanodisc antenna were fabricated by evaporating a gold film on a fused silica substrate.
The film thickness exceeded \nm{200} to exclude effects from the substrate side for this sample.
The structures were milled into the gold film using Focused Ion Beam Milling (FIB) with gallium ion irradiation.
The width of the milled grooves was \nm{80}, the central disc had a diameter of \nm{180} and the period of the grating was \nm{780}.

\begin{figure}
	\centering
	\includegraphics{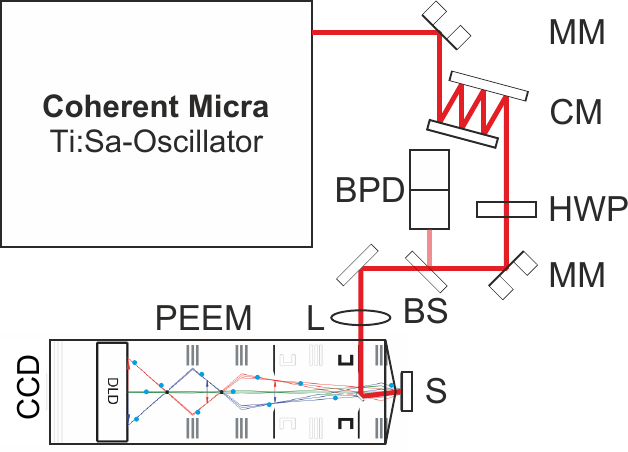}
	\caption{Sketch of the PEEM setup. A polarized \fs{120} pulse is focused on the sample S with a lens L (focal length \SI{250}{\milli\metre}). The polarization is controlled by a half-wave plate (HWP). The beam was stabilized using the beam stabilization system \textsc{Aligna} (TEM Messtechnik Germany), allowing for stability during long-term measurements. A part of the beamline was passed to a beam position detector (BPD) behind a beam splitter (BS) for this purpose and the feedback was used to control the motorized mirrors (MM). The chirped mirrors (CM) compressed the pulse to the final duration of \fs{30}.}
	\label{fig:peem}
\end{figure}

The experimental setup is shown in Fig.~\ref{fig:peem}.
As laser source we used a mode-locked Ti:Sa oscillator (\textsc{Coherent} Micra), which provides an average output power of \SI{100}{\milli\watt} at the sample position at a repetition rate of \SI{80}{\mega\hertz}, resulting in a pulse energy of approximately \SI{1}{\nano\joule}.
The duration of the ultrashort pulses was \fs{30}, the central wavelength was \nm{800} and the spectral bandwidth \nm{60}.

The frequency of the laser oscillator corresponds to a photon energy of \SI{1.55}{\electronvolt}.
The work function for gold evaporated in our facilities is in the order of \SI{4.6}{\electronvolt}.
This means that photoemission of electrons is just possible in a nonlinear 3-photon process which requires enough intensity in the optical field in order to be observed.

We used a PEEM from Focus GmbH (Germany) to image the emitted photoelectrons.
The illumination angle of incidence was \SI{4}{\degree} with respect to the sample normal.
A wide Gaussian beam was imaged onto the sample using a $ f=\SI{250}{\milli\metre} $ lens, resembling a quasi-plane-wave illumination on the sample.
Photoexcited electrons emitted from the sample are accelerated by a static extractor voltage up to \SI{16}{\kilo\volt} and imaged by an electrostatic lens system which provides up to \nm{20} lateral resolution depending on the field of view.
The amount of emitted electrons can be expected to be small enough in order to not influence each other.
The total exposure time was about \SI{15}{\minute} for the ring-enhanced structure.
This was optimized to give the best signal-to-noise ratio in the experiments.

\subsection{Results and interpretation of the PEEM measurements}
\label{PEEM}

\begin{figure}
	\centering
	\includegraphics[height=5.5cm]{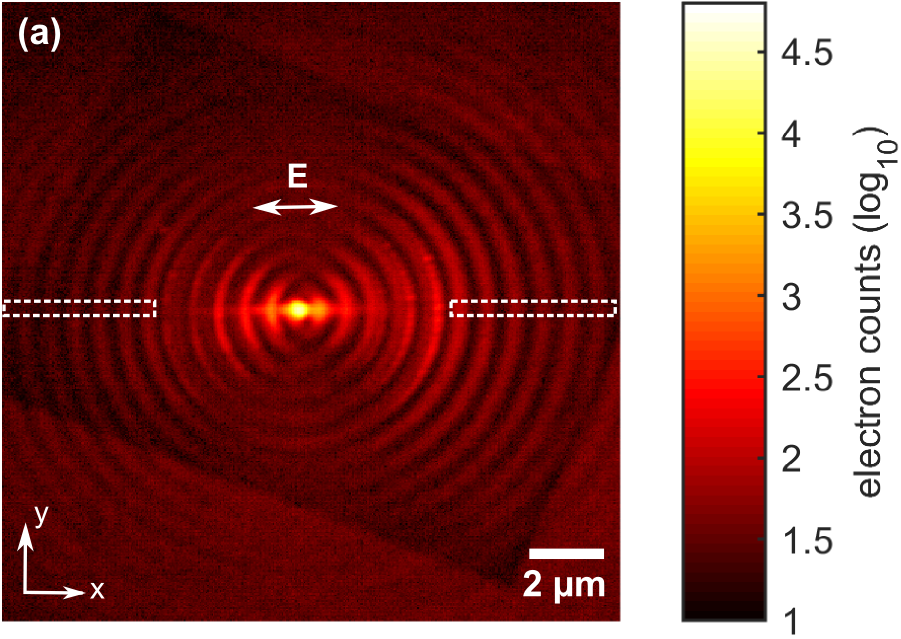} \\[3ex]
	\includegraphics[height=5.5cm]{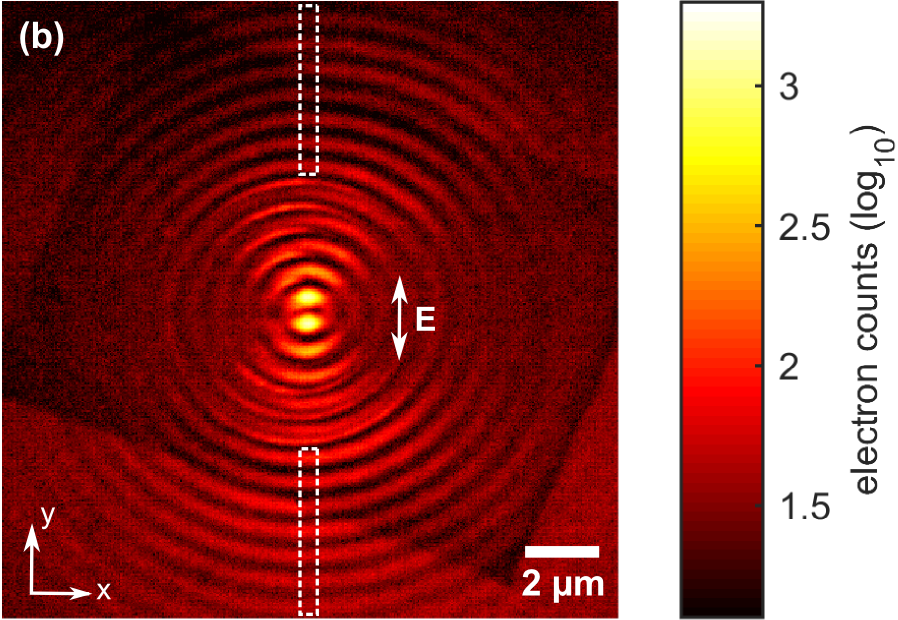}
	\caption{Logarithmic plot of the PEEM yield for the ring-enhanced sample in (a) TM and (b) TE polarization. Both polarizations show noticeable yield but only TM polarization excites the central disc. The dashed lines denote the areas from which the data for the detailed analysis of outward-propagating Hankel plasmons visible in the outer parts was extracted (Sec.~\ref{hankel}).}
	\label{fig_peem_overviews}
\end{figure}

The experimental findings for the ring-enhanced structure in TE and TM polarization are shown in Fig.~\ref{fig_peem_overviews}.
We observe first of all a dark area around the sample structure, which is caused by the SEM inspection during fabrication.
The scanning electron beam provides the energy to permanently adsorb chemicals (most likely carbon) which are present.
Since PEEM is extremely sensitive to surface conditions, even a monolayer of adsorbed material can alter the effective work function.
In our case, it is slightly increased in the scanned area which explains the darker region in the center.
The outer region represents parts which were not scanned by the electron beam.

\begin{figure}
	\centering
	\includegraphics[height=5.5cm]{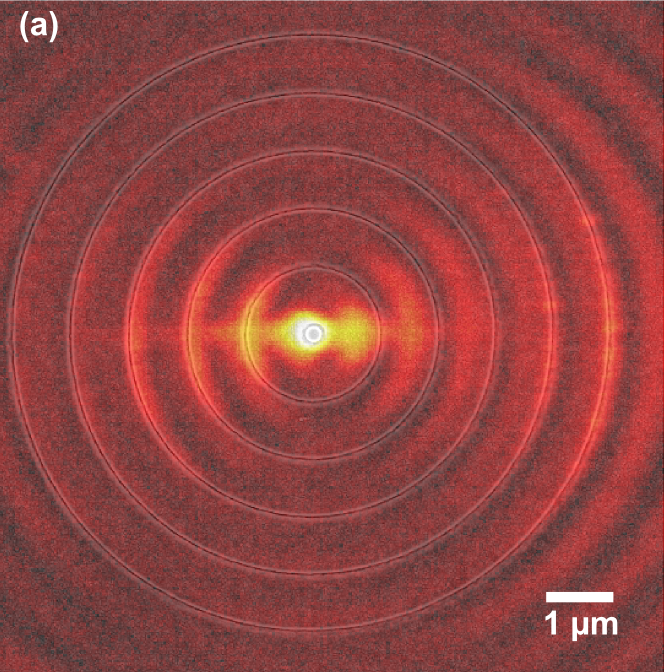} \\[3ex]
	\includegraphics[height=5.5cm]{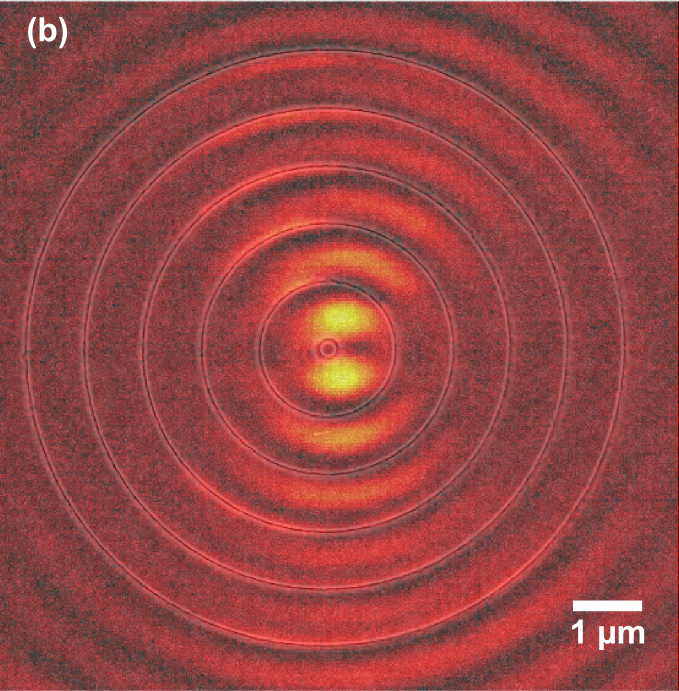}
	\caption{PEEM images overlayed with the SEM image for (a) TM and (b) TE polarization. A clear qualitative difference in the field localization and excitation of the central nanodisc is observed as well as details of the localization of the photoemission.}
	\label{fig_peem_overlays}
\end{figure}

Although the effect seems disturbing it actually increases the signal-to-noise ratio of our measurements and provides the possibility to accurately overlay PEEM images with the SEM image.
This allows us to create the overlay plots shown in Fig.~\ref{fig_peem_overlays}, where the origin of the PEEM yield can be studied in great detail with high spatial resolution.
We estimate the positioning accuracy of the overlay to be about \nm{60}.

Photoemission is known to be very sensitive to the presence of an electric field component perpendicular to the surface \cite{Fecher2002}.
In the case of TM illumination, we thus find a strong excitation of the central disc.
Both polarizations were ensured to carry the same power and were measured using the same exposure time.
Apart from the strong resonance in the center, we also find a weaker lobe-like enhancement in the horizontal direction, which was the plane of incidence.
This is consistent with Hankel plasmon theory, which predicts that the excitation intensity drops as $ \cos^2(\varphi) $ for TM modes, where $ \varphi $ is the angle to the horizontal direction.
Consequently, no excitation of HSPPs is observed in the vertical direction.
It should also be noticed that the photoelectron yield is slightly increased at the positions of the rings.
Although there cannot be electron emission from the gaps, sharp edges lead to a higher probability for electron emission due to confinement effects.
This manifests itself in the increased yield at the ring positions since the gaps are too narrow to be accurately resolved.

We were able to measure a weaker but noticeable yield also for TE polarization.
The reason for this is, that the excited HSPPs themselves provide for the necessary perpendicular electric field component.
This argumentation is supported by the fact that we measured a much lower yield for the bare nanodisc sample in TE polarization as well, where the excitation of HSPPs will be shown to be only very weak in Section~\ref{bare_disc}.
In the ring-enhanced case, a clear excitation along the vertical image direction is visible.
For this polarization, theory predicts a $ \sin^2(\varphi) $ dependence of the intensity, which is in agreement with our experiments.
However, we find that the picture has clear qualitative differences to the TM case.
The central disc was not excited.
Instead, a larger dipolar Hankel mode in vertical direction forms on the area between the disc and the first ring.
We again observe a lobe-like spatial structure in polarization direction and an increased yield from the ring positions, similar to the TM case.

\subsection{Comparison to FDTD data}
\begin{figure*}
	\centering
	\includegraphics{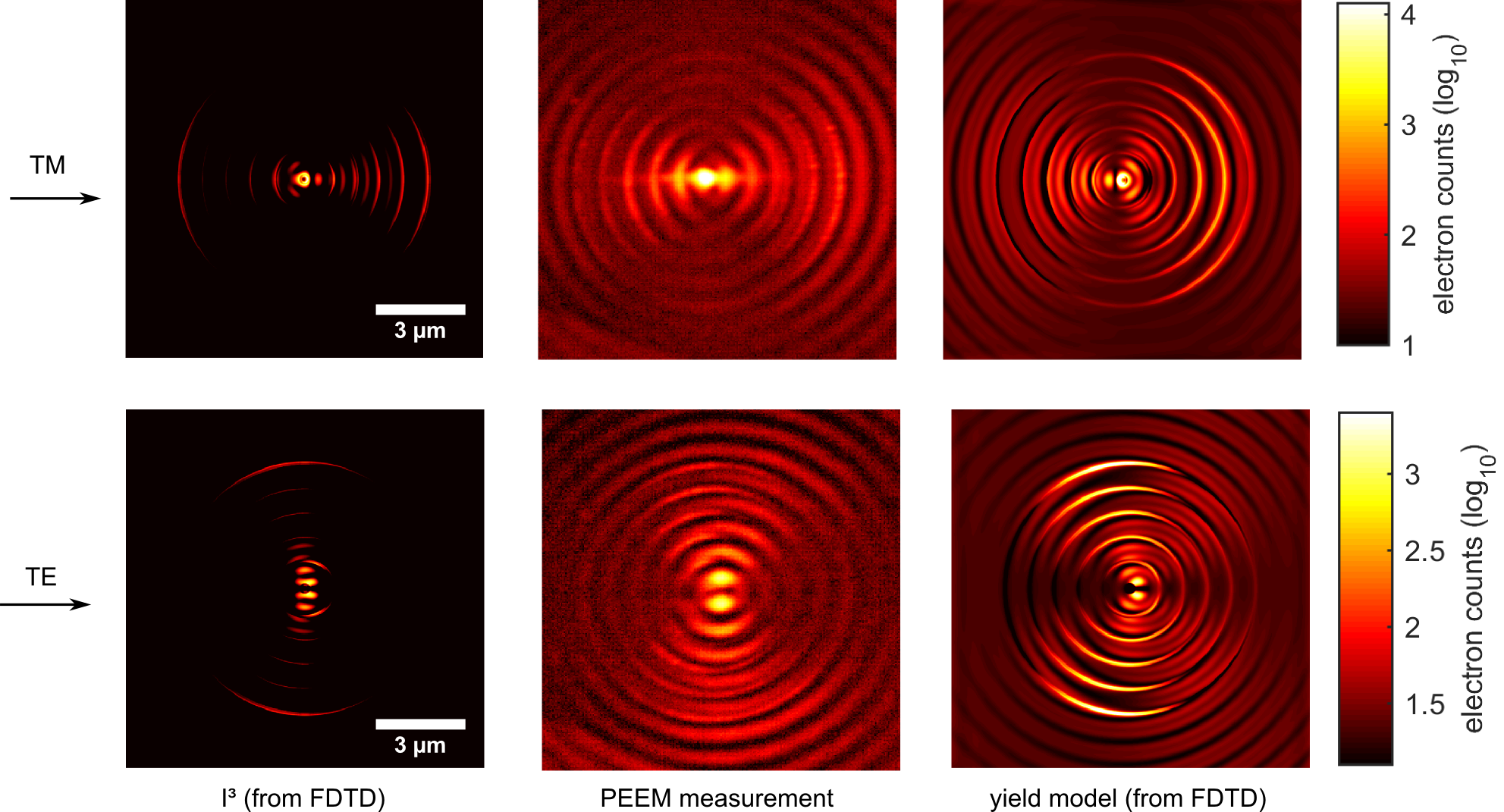}
	\caption{Comparison between the third power of the intensity from FDTD data, the actual PEEM measurement and the yield computation from the model described in the text. The colorbars apply to all plots in the same row. The yield model is in very good agreement with the measurement, whereas the intensity data does not give the correct background levels.}
	\label{fig_peem_comparison}
\end{figure*}

We compared the experimental PEEM images to rigorous simulations using the finite-difference time-domain method (FDTD).
Full 3D simulations were employed under \SI{4}{\degree} incidence using the freely available FDTD software meep \cite{Oskooi2010}.
The simulation time was chosen long enough for the field to reach the steady-state.
The results are shown in Fig.~\ref{fig_peem_comparison}.

The left panel shows the intensity calculated close to the sample surface for both polarizations.
It becomes evident that the qualitative difference we observed between the two polarization directions is inherent to the investigated structure.
The TM simulation confirms that the central disc is the location of the highest intensity in this case whereas it is not excited in the TE case.
The symmetry of the system is broken by the \SI{4}{\degree} incidence angle while the additional rings increase the angular sensitivity of the antenna.

Although the FDTD intensity results confirm in principle our PEEM measurement data, the images look quantitatively different.
The background level within the ring area is much higher in the measurement and no pronounced two-lobe structure is seen for the TE case.
To explain this discrepancy, one must not forget that the photoelectron yield is not a direct measure for the electromagnetic fields.
An accurate modeling is needed to connect the two quantities.

We therefore employ a model which connects the photoelectron yield to the temporal integral of the sixth power of the instantaneous electric field.
This is facilitated by the fact that the probability for three-photon photoemission should be proportional to the absolute value squared of the matrix element for three photon absorption, following a nonlinear generalization of Fermi's golden rule \cite{Boyd2008}.
This matrix element involves the interaction Hamiltonian $ \vec p \cdot \vec E $, where $ \vec p $ is an atomic dipolar moment and thus leaves us with the overall dependence on the electric field to the sixth power in the 3-photon case.
This approach differs from using the third power of the local intensity, since intensity is defined as a time-average of the Poynting vector \cite{Born1991}
\begin{equation}
I = \biggl|\frac 1 2 \Re \Bigl[\vec E(\vec r, \omega) \times \vec H^*(\vec r, \omega)\Bigr]\biggr| ,
\end{equation}
although the principle overall power dependence $ Y \propto I^3 $ stays intact and differences occur  just in the weights of certain polarization components contributing locally to the PEEM yield image.

The total value of the instantaneous electric field involves polarization contributions perpendicular and parallel to the sample surface
\begin{equation}
\left|\vec E(\vec r, t) \right| 
= \left[ E_\perp^2(\vec r, t) + E_\parallel^2(\vec r, t) \right]^\frac{1}{2} .
\end{equation}
We therefore arrive at our used model formula for the 3-photon PEEM yield
\begin{equation} \label{eq:yieldmodel}
Y(\vec r) \propto \int \left|\vec E(\vec r, t ) \right|^{6}  \,{\rm d}t 
= \int  \left[E_\perp^2(\vec r, t) + E_\parallel^2(\vec r, t) \right]^3 \,{\rm d}t .
\end{equation}

For the Hankel plasmon, both polarization contributions are present in form of a perpendicular and radial component and approximately $ \pi/2 $ out of phase.
The illuminating plane wave either contains only a parallel component (TE case), or both components in phase (TM case) which are superimposed during the duration of the pulse.
Since we use just a single ultrashort \fs{30} pulse to perform the experiment, HSPPs are excited from the early parts of the pulse and interfere with later parts in a spatio-temporal coherent manner.
This creates the interference patterns typical for multiphoton PEEM experiments.

To take this into account, we modeled the PEEM yield by coherently superimposing the nanoantenna near-field extracted from the FDTD data with the corresponding illuminating field components, allowing the amplitude and total phase of the latter as a free fit parameter.

The last panel of Fig.~\ref{fig_peem_comparison} shows the result of the modeling.
The same fit parameters were used for both polarizations.
A remarkable agreement with the measurement data is achieved.
The TE case shows the correct two-lobe structure now.
Additionally, background levels and even finer details are accurately reproduced qualitatively and quantitatively.

A peculiar difference is that the model overestimates the yield from the positions of the rings.
The reason for this is the plasmonic field enhancement within small gaps, which is present in the FDTD data.
Since the gaps are not filled with metal, no enhancement of the photoelectron yield can occur at these positions apart from the sharp edges as discussed earlier. 

\subsection{Analysis of propagating Hankel plasmons}
\label{hankel}
In addition to the antenna resonance, we observe outward-propagating HSPPs on the surrounding gold film in both polarization cases.
They can be analyzed in detail since a full analytical theory is readily available \cite{Filter2012}.
We confirmed that the measured HSPPs obeyed the predicted angular dependence of the intensity as $ \cos^2(\varphi) $ and $ \sin^2(\varphi) $, respectively (not shown).

\begin{figure}
	\centering
	\includegraphics[width=8.3cm]{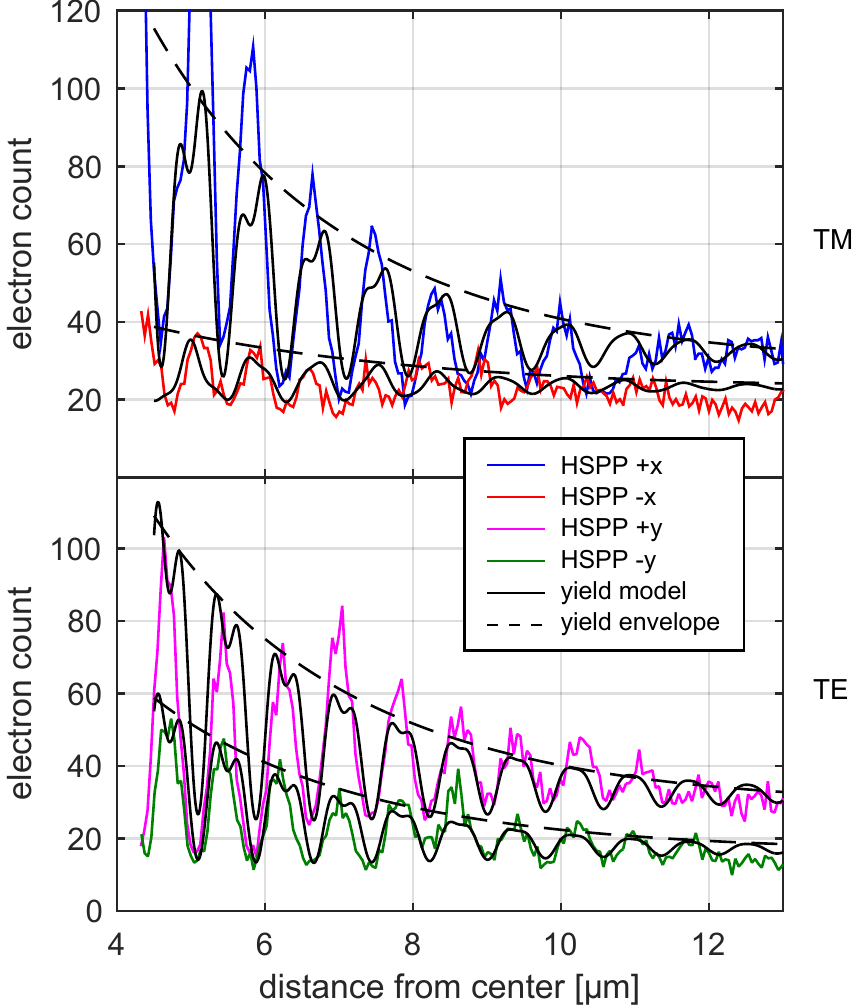}
	\caption{Detailed analysis of outward propagating Hankel plasmons from the regions marked in Fig.\ref{fig_peem_overviews} and comparison to the analytical model. The yield envelope contains information about the temporal structure of the pulse as well as the spatial decay of Hankel plasmons.}
	\label{fig_Hankelmodel}
\end{figure}

To further investigate their properties, we extracted a cut through the horizontal (TM) or vertical direction (TE), shown as dashed areas in Fig.~\ref{fig_peem_overviews}.
We plot the results in Fig.~\ref{fig_Hankelmodel}.
In the TE case, the HSPPs propagate perpendicular to the plane of the incident pulse.
The measured oscillation period therefore matches exactly the plasmon wavelength $ \lambda\tief{hspp} $.
For TM polarization, the HSPP pulse propagates parallel or antiparallel to the excitation pulse direction, respectively.
The wavenumber of the oscillation is thus altered by $ \pm k_0 \sin \theta $ in this case.
However, this effect is only minimal, since $ \sin(\SI{4}{\degree}) \approx 0.07 $.

The attenuation of the oscillations is governed by two terms.
The first contribution comes from the physics of the Hankel plasmon itself.
As radial solutions, they drop with a factor $ [\kspp r]^{- \frac 1 2} $ quicker than plane surface plasmon polaritons.
The second factor originates from the limited illumination time during the ultrashort pulse.
If we assume a Gaussian envelope, the temporal width $ \tau $ of the pulse translates into a spatial envelope for the excitation which leaves us with the total dependence of the envelope
\begin{equation} \label{eq:envelope}
Y\tief{env}(r) \propto 
\Biggl\{
\biggl[ \Re(\kspp) r \biggr]^{-\frac 1 2}
\exp\left[- \left( \frac{n_g r}{\tau c} \right)^2 \right] + {\rm const.}
\Biggr\}^6 ,
\end{equation}
where $ n_g $ is the group index of the plasmon pulse and $ r $ is the distance to the origin.
The vacuum speed of light is denoted by $ c $.
From the dispersion relation of HSPPs at \nm{800} free-space wavelength, we calculate $ n_g = 1.08 $.
The term $ \tau c $ is approximately \um{9} and represents a characteristic length scale for the experiments.
The analytical model fits the experimental data excellently.
This also opens the possibility to check or even measure the pulse length of the excitation from the PEEM images.

We emphasize that equation \eqref{eq:envelope} does not contain a contribution which is inherent to the intrinsic ohmic metal loss, \ie a term proportional to $ \exp[-\Im(\kspp) r] $.
This term can be neglected since 
\begin{equation}
\frac {\Im(\kspp)} {\Re(\kspp)} \approx 10^{-3}
\end{equation}
and does not contribute noticeably over the distances relevant in our experiment.

We use the yield model described above to fit the experimental data.
The total electric field is a superposition of the HSPP and the illuminating plane wave with  excitation amplitudes $ A $ and $ B $
\begin{equation}
\vec E(\vec r, t) = A \cdot \vec E\tief{hspp}(\vec r, t) + B \cdot \vec E\tief{pw}(\vec r, t) ,
\end{equation}
where both terms contain a moving Gaussian pulse envelope
\begin{align}
\vec E\tief{pw}(\vec r, t) &= \vec{\tilde{E}\tief{pw}}(\vec r, t) 
\cdot \exp\left[- \frac{(t - \sin\theta x / c)^2}{\tau^2}\right] \\
\vec E\tief{hspp}(\vec r, t) &= \vec{\tilde{E}\tief{hspp}}(\vec r, t) 
\cdot \exp\left[- \frac{(t - n_g r / c)^2}{\tau^2}\right]
\end{align}
in addition to their steady-state solutions
\begin{align}
\vec{\tilde{E}\tief{pw}}(\vec r, t) &=
\frac {1} 2 \biggl[ \vec E_0 \e^{ik_0 \sin\theta x - i\omega t} + ~{\rm c.c.} \biggr] \\
\vec{\tilde{E}\tief{hspp}}(\vec r, t) &= \frac 1 2 \biggl[ \vec E\tief{hspp}(\vec r, \omega) \e^{-i\omega t} + ~{\rm c.c.}\biggr] .
\end{align}
$ \vec E_0 $ denotes the polarization vector in the TE or TM case
\begin{equation}
\vec E_0 = 
\begin{cases}
[0, 1, 0] & \text{(TE)} \\
[\cos\theta, 0 , \sin\theta] & \text{(TM)}
\end{cases} 
\end{equation}
and  $ \vec E\tief{hspp}(\vec r, \omega) $ the analytical HSPP solution given in eq.~\eqref{eq:hspp}.

The yield model \eqref{eq:yieldmodel} is used with the above field expressions to fit the excitation amplitudes to the experimental data in Fig.~\ref{fig_Hankelmodel}.
The results are in very good agreement with the experimental findings.
Slightly different fitting values had to be used in the different cases since the background levels far from the center were not ideally equal at all positions.
We believe that a slight spatial inhomogeneity of the illumination at the sample position is the reason for this since it is a wide Gaussian beam in reality and not an ideal plane wave.
Some deviations from the measurement also occur for smaller distances from the center which happens since the wave originates from different rings in the experiment and not just from an ideal point source.
The approximation introduced for HSPPs in \eqref{eq:hspp} therefore does not fully hold close to the outermost ring.
Nevertheless, we were able to show that the PEEM images contain rich spatio-temporal information  which allows a detailed analysis and understanding of the underlying physics of the structure.

%
%
%
%
\subsection{Comparison with measurements on the bare nanodisc}
\label{bare_disc}
\begin{figure}
	\centering
	\includegraphics[width=8cm]{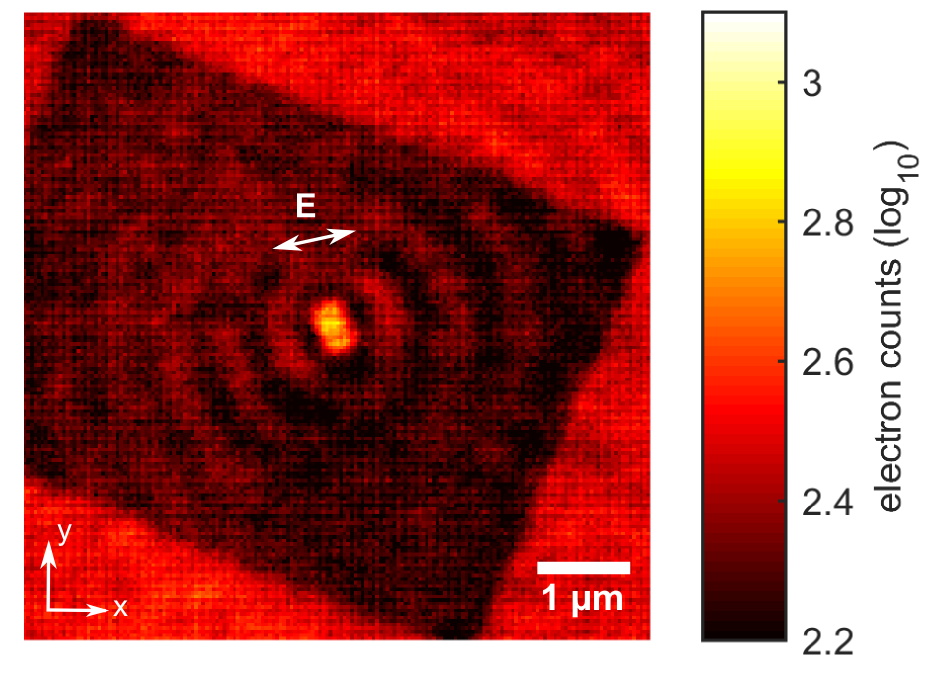}
	\caption{Logarithmic plot of the PEEM yield for the bare nanodisc sample in TM polarization. An enhancement of photoemission due to the plasmonic resonance is observed at the position of the disc which is weaker than in the ring-enhanced case.}
	\label{fig_peem_single_disc}
	\end{figure}

The results for the ring-enhanced structure are compared to the measurements on the bare nanodisc in order to quantitatively verify the enhancement mechanism of the antenna.
The total exposure time had to be increased to $ \approx \SI{2}{\hour} $ in order to get a significant signal.
Fig.~\ref{fig_peem_single_disc} shows the experimental result for the bare plasmonic nanodisc.
The polarization of the incident light was TM in this case, the electrical field vector being in the plane of incidence.
The yield of photoelectrons at the position of the nanodisc was found to be much stronger than on the surrounding planar gold film, but much weaker than in the ring-enhanced case.
This is caused by the localized quasi-static plasmonic resonance of the nanodisc, in accordance with our theory.
The incident plane was not exactly horizontal in Fig.~\ref{fig_peem_single_disc} which can be seen from the weak excitation of HSPPs excited by the single groove and propagating away from the disc.
However, the excitation strength is so low that we could not observe noticeable yield for TE polarization using the same exposure time (not shown). In this case, the HSPP is the only source of a normal electric field component.

If we compare the maximum values for $ Y^{1/3} $ in the TM case and take into account the different exposure times, we find an about 6-fold enhancement of the near-field intensity at the nanodisc by adding the ring grating.


\section{Comparison with SNOM measurements}
\label{SNOM}
\begin{figure*}
	\centering
	\includegraphics{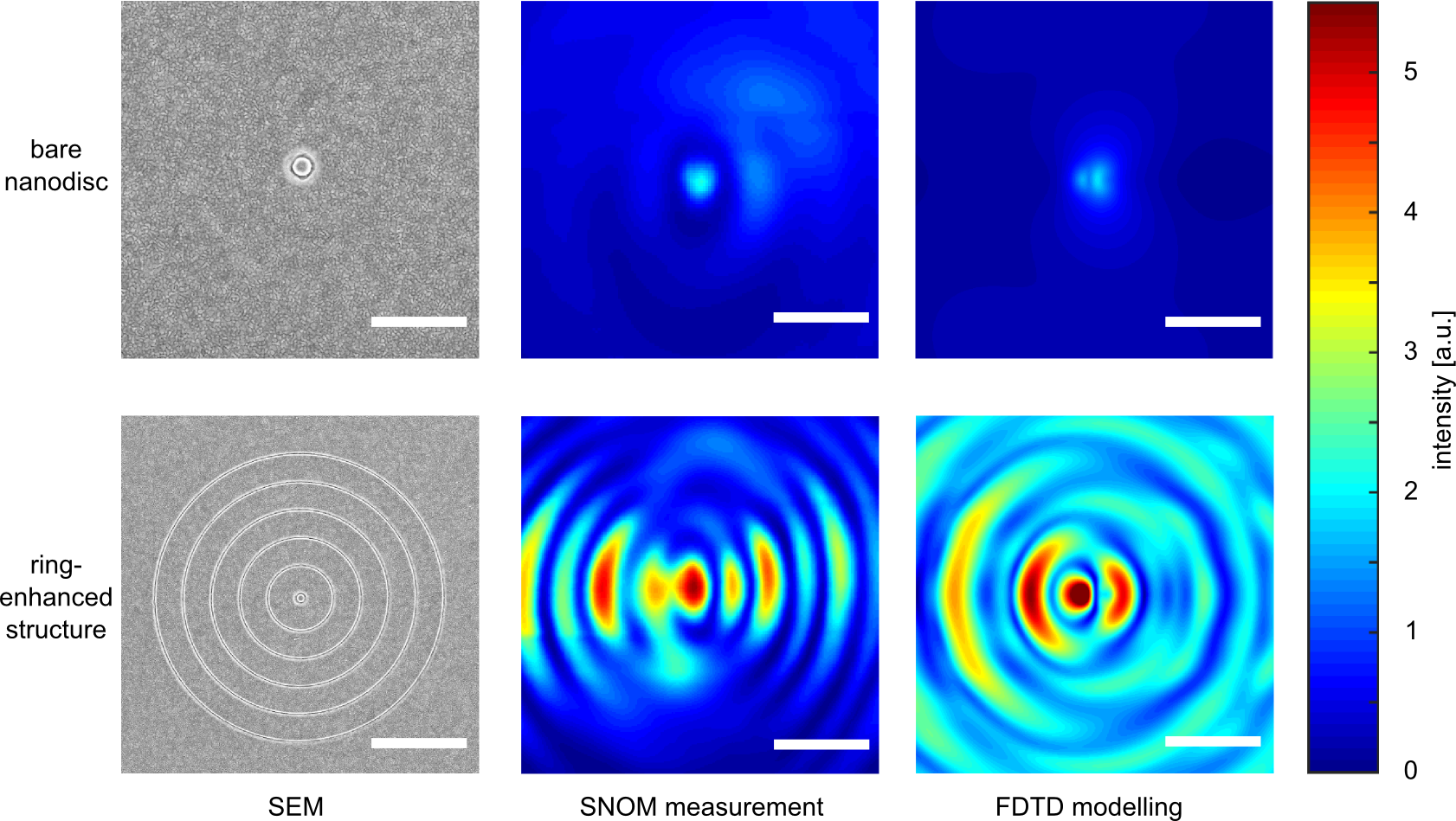}
	\caption{Comparison of SNOM measurement results and FDTD modeling for the bare nanodisc sample and the ring-enhanced structure. Measurement and model verify the enhancement of the near-field intensity. The scalebar is \um{2} long in each image. The spatial resolution is limited to \nm{180}, which was the apex size of the aperture SNOM tip.}
	\label{fig_snom}
\end{figure*}
The PEEM results are compared to previous measurements \cite{Qi2015} on the same structure using scanning near-field optical microscopy (SNOM) as established and accepted experimental method for nanophotonics.
We had to adapt the initially described sample design slightly in order to account for the experimental circumstances.
The experiments were performed with a diode laser of \nm{785} wavelength which illuminated the structure from the substrate side.
A SNOM tip with a circular aperture was used to scan the sample from the top.
The thickness of the gold film had to be decreased to \nm{50} in order to allow sufficient coupling of light to the upper side.
This results in a certain interaction between plasmons at the substrate and cladding interfaces.
It was shown in a previous work, that this interaction leads to a slightly altered value for \kspp, but does not influence the basic functionality of the structure much, neither qualitatively nor quantitatively if we regard the field profiles and enhancement factor, which are our values of interest here \cite{Qi2015}.

A wide and polarized Gaussian beam was imaged at the sample position so that the excitation resembles a plane wave.
The aperture tip was gold coated and light was collected into a single mode optical fiber which guided the signal to a photodetector.

Fig.~\ref{fig_snom} shows a summary of the SNOM results.

We started by investigating the single disc first.
A clear excitation of a localized plasmonic resonance at the disc position was observed, accompanied by a small amount of background light from direct transmission of the illumination.
The structure surrounded by the rings showed a much more complex behavior.
We found pronounced side lobes which we attribute to the excitation of Hankel plasmons by the grating.
They appear only in horizontal direction in the images which is consistent with the incident polarization direction and the PEEM findings.
We also observe an enhancement of the field strength at the central disc.
The near-field intensity was again approximately 6 times higher for the ring-enhanced structure than for the bare nanodisc.
This shows that the PEEM measurements lead to quantitatively comparable results.

The SNOM data was also compared to rigorous FDTD simulations.
We had to account for certain experimental circumstances by processing the raw simulation data.
The tip is sensitive to electric field components perpendicular to the tip axis and approaches the sample under an angle of \SI{30}{\degree} \cite{Klein2014}.
Furthermore, the apex size of approximately \nm{180} limits the achievable resolution. 
This means that all images get slightly blurred as if they were taken further away from the sample.
We took that into account by projecting the FDTD data on a plane inclined by \SI{30}{\degree} with respect to the surface normal and used an extraction distance of \nm{100} above the structure.
The results are shown in Fig.~\ref{fig_snom} and show a good agreement with the experimental findings.
This demonstrates the superior possibilities of PEEM over SNOM with respect to resolution.
In both cases, accurate modeling of the underlying physics is necessary to interpret the resulting images and compare to, \textit{e.g.}, rigorous simulations or analytical data.

\section{Conclusion}
\label{conclusion}

We investigated a circular optical nanoantenna under ultrafast femtosecond laser irradiation be means of multiphoton photoemission electron microscopy.
Positions at which the electromagnetic field got enhanced by the constructive interference of Hankel plasmons were found to lead to a much higher yield in photoelectrons.
The method thus provides information about the spatial structure of the electromagnetic field with high accuracy on ultrashort timescales.

For our investigations, we used an antenna design composed of a plasmonic nanodisc, surrounded by rings which function as a grating.
It acts as a Hankel plasmon coupler on the gold film used for the experimental realization of the device.
Inward propagating Hankel plasmons are resonantly excited by the illumination and drive a localized resonance at the central disc of the structure, where the grating confines the energy by a Bragg resonance. This enhances the near-field intensity by a factor of $ \approx 6 $ compared to the localized plasmonic resonance of the disc alone.
The grating also excites outward propagating Hankel plasmons which were observed in the PEEM and shown to contain information about the propagating plasmons themself, as well as the ultrashort excitation of the structure by a femtosecond pulse.

We compared our findings to measurements of the same antenna design using scanning near-field optical microscopy (SNOM).
While this method has the advantage of directly mapping the electromagnetic near-field, it was shown that the physical presence of a probe has disadvantages regarding the experimental configuration and achievable resolution.
Moreover, the measurement of ultrashort electromagnetic phenomena is very difficult due to the scanning nature of the method.
The achieved PEEM images had a much better spatial resolution than the SNOM images and do not require any disturbing probe or sample scanning which leaves much freedom to the possible excitation scenarios.
However, it was shown that accurate physical modeling is necessary to relate the acquired PEEM images to the electromagnetic field since the process is of nonlinear nature.

Since the photon energy of the used \ev{1.55} radiation is much lower than the workfunction of gold, a 3-photon process was necessary to excite photoelectrons.
The photoelectron yield was modeled as the temporal integral of the instantaneous yield, which was shown to depend on the total electric field to the sixth power in our case.
This field was composed of an electromagnetic interference between both the plasmon fields and the illumination itself.
Rigorous electromagnetic simulations were performed using the finite-difference time-domain method (FDTD).
An analytical theory was employed as well to describe the outward propagating Hankel plasmons.
Both were found to be in very good agreement with the experimental images when we apply our yield model to the theoretical and numerical data.

Nevertheless, more work has to be done in the future to clarify the exact role of the local polarization as a function of time for the acquisition of the multiphoton PEEM images.
This includes the implementation of a pump-probe configuration in our setup to probe the complete spatiotemporal dynamics of the structure.

\section{Acknowledgements}
We thank N. Asger Mortensen (DTU Denmark) for stimulating discussions and Focus GmbH (Germany) for the sketch of the PEEM in Fig.~\ref{fig:peem}.
Funding is acknowledged by Deutsche Forschungsgemeinschaft (DFG SPP 1391 Ultrafast Nanooptics), the Thu\-ringian Ministry for Economy, Science and Digital Society (TMWWDG Pro-Ex\-zel\-lenz program), and the Carl Zeiss foundation.
%
%

\end{document}